\documentstyle[12pt,epsfig,preprint,aps]{revtex}

\def \to {\rightarrow}
\def \beq {\begin{equation}}
\def \eeq {\end{equation}}
\def \ba {\begin{eqnarray}}
\def \ea {\end{eqnarray}}
\def \jpsi {J/\psi}
\def \mtss {\langle 0 |{\cal O } ^{J/\psi}_{1} [ ^3S_1 ]| 0 \rangle}
\def \mtso {\langle 0 |{\cal O } ^{J/\psi}_{8} [ ^3S_1 ]| 0 \rangle}
\def \moso {\langle 0 |{\cal O } ^{\jpsi}_{8} [ ^1S_0 ]| 0 \rangle}
\def \mtpo {\langle 0 |{\cal O } ^{\jpsi}_{8} [ ^3P_0 ]| 0 \rangle}
\def \mtpj {\langle 0 |{\cal O } ^{\jpsi}_{8} [ ^3P_J ]| 0 \rangle}

\begin{document}
\draft
\title{ $ \jpsi + jet $ diffractive production in the direct photon process
at HERA }
\author{Jia-Sheng Xu}
\address{Department of Physics, Peking University, Beijing 100871, China}
\author{Hong-An Peng}
\address{ China Center of Advance Science and
Technology (World Laboratory), Beijing 100080, China \\
and Department of Physics, Peking University, Beijing 100871,
 China }
\author{Zhan-Yuan Yan, and Zhen-Min He}
\address{Department of Physics, Hebei Teacher's University,
Shijiazhuang  050016, China}

\maketitle

\begin{abstract}
We present a study of $\jpsi + jet$ diffractive production in the direct
photon process at  HERA based on the factorization theorem for lepton-induced
hard diffractive scattering and the factorization formalism of
the nonrelativistic QCD (NRQCD) for
quarkonia production. Using the diffractive gluon distribution function
extracted from HERA data on diffractive deep inelastic scattering and
diffractive dijet photon production,  we show that this process
can be studied at HERA with present integrated luminosity,
and can give valuable insights in the color-octet mechanism for heavy
quarkonia production.
\vskip 3mm
\end{abstract}

\vfill\eject\pagestyle{plain}\setcounter{page}{1}


\section{Introduction}

Diffractive scattering was imported into high energy strong interaction
to interpret the $ t $ distribution of elastic hadron-hadron scattering
at small $ |t| $ region \cite{perl}. In the framework of the Regge theory,
diffractive scattering is assumed to take place through the exchange of an
object with the quantum numbers of the vacuum (the Pomeron) \cite{collins},
Although, the Pomeron plays a particular and very important role in soft
processes in hadron-hadron collisions, the true nature of it and its
interaction with hadrons remain a mystery.

\par
Due to the nature of color singlet exchange, the diffractive
scattering events  are characterized by a large rapidity gap, a region in
rapidity devoiding of hadronic energy flow.
Events containing rapidity gap and jets were first observed by  
UA8 Collaboration at CERN \cite{ua8}, opening the field of hard
diffraction. The hard diffraction is also observed and studied 
by  various experiments \cite{gap,zeus,h1,cdfw,cdfjet}.
Although it is still difficult to understand the soft diffractive
scattering in the framework of QCD, much progresses in understanding
the nature of hard diffraction are made in the light of great theoritical
and experimental efforts \cite{diff-rev}.
As an attempt of understanding hard diffraction on the parton level,
Ingelman and Schlein \cite{ingelman} assumed that the Pomeron, similar to
the nucleon, is composed of partons, mainly of gluons, and the hard
diffractive processes can probe the partonic structure of the Pomeron.
In fact, the Ingelman-Schlein model for hard diffraction
is based on the assumption of hard diffractive scattering factorization
together with  Regge factorization for the Pomeron exchange.  
But, comparison the total $pp({\bar p})$ single diffraction cross section
data \cite{sd-ex} with predictions based on the Regge factorization for the
Pomeron exchange indicates the breakdown of the Regge
factorization \cite{goulianos}.
Furthermore, as pointed by Ingelman in \cite{diff-rev}, there are
conceptual and theoretical problems with the Ingelmal-Schlein
model. First, the Pomeron is not a real state, it is only a virtual
exchanged spacelike object. the concept of a structure function is then
not well defined. In fact, the factorization of diffractive structure
function into a Pomeron flux and a pomeron structure function cannot be
uniquely defined since only the product is an observable quantity.
Second, the Pomeron-proton interaction is soft, its space-time scale is
larger compared to that of the hard interaction, so
the Pomeron can then not be considered as decoupled from the proton, and
in particular, is not a separate part of the QCD evolution in the proton.

\par
For the lepton induced hard diffractive scattering processes,
such as diffractive deep inelastic scattering (DDIS) and diffractive
direct photoproduction of jets, there is indeed a factorization theorem
which  has been proven by Collins \cite{jccollins} recently.
It is the same as factorization for inclusive hard processed, except
that parton densities are replaced by diffractive parton
densities \cite{kunszt,berera}.
The diffractive parton densities (the primary non-perturbative quantities in
hard diffraction) obey exactly the same DGLAP evolution equations as
ordinary parton densities, and have been extracted from 
HERA data on DDIS and on diffractive photonproduction of jets \cite{alvero}.
This factorization theorem also establishes  the universality of the
diffractive parton distrubutions for those processes to which the theorem
applies, hence justifies, from fundamental principles,  the analysis of
ZEUS and H1 Collaboration \cite{h1,zeus} on hard diffraction, provided the
term `` Pomeron ''  used in these analyses is as a label for a particular
kind of parametrization for diffractive parton
densities, and an indication of the vacuum quantum numbers to be exchanged.
In the light of hard diffractive factorization proven by
Collins \cite{jccollins}, the Ingelman-Schlein model for hard diffraction
assumed further the factorization of diffractive parton densities into
a universal Pomeron flux and a Pomeron parton densities,
but such  further factorization is unjustifiable.

\par
In this paper, we present a study of  another hard diffractive process, 
$\jpsi + jet$  diffractive production in the direct photon process at HERA
\beq
\label{proc}
\gamma (q) + p(p_1) \to p(p_{1}^{\prime}) + \jpsi (P) + jet + X ,
\eeq

\noindent
where $ q, p_1, p_{1}^{\prime}$, and $P$ are the momenta of
incoming photon, proton, outgoing proton, and $\jpsi$ respectively.
We base on the factorization theorem for lepton-induced
hard diffractive scattering and the factorization formalism of
the nonrelativistic QCD (NRQCD) for
quarkonia production. Using the diffractive gluon distribution function
extracted from HERA data on DDIS and diffractive dijet photon production,
we show that this process can give valuable insights in the color-octet
mechanism for heavy quarkonia production.
This process was briefly discussed in \cite{jung} where the relativistic
corrections to photoproduction of $\jpsi$ were studied in the so-called
color-singlet model (CSM)\cite{csm}.

\par
Our paper is organized as follows. We describe in detail our calculation
scheme in Sec.II, in which, a brief introduction of the heavy quarkonium
production mechanism, the hard subprocesses for $\jpsi + jet $ production
and a summary of the kinematics related to  the $\jpsi$'s $z$ and $P_T$
distributiones are included.  Our results and discussions are
given in Sec. III.

\section{Calculating Scheme}
Based on the factorization theorem for lepton-induced
hard diffractive scattering and the factorization formalism of NRQCD for
heavy quarkonia production, $\jpsi + jet $ diffractive production process 
is shown in Fig.1 . This process consists two steps.
First, the almost point-like $c{\bar c}$ pair with large $P_T$ is produced
in the hard subprocesses, then $ \jpsi $ particle is produced from this
point-like $c{\bar c}$ pair via soft interactions.

\subsection{Heavy quarkonium production mechnism}
Heavy quarkonium production in various processes has been the focus  of
much experimental and theoretical attention during the past few
years \cite{review}. This is mainly due to the observation of large
disrepancies between experiment measurements of prompt and direct $\jpsi$
production and $\psi^{\prime} $ production at large $P_T$ at the Collider
Detector Facility (CDF) at the Fermilab Tevatron and the calculations
based on  CMS \cite{cdfjpsi}.

\par
The first major conceptual advance in heavy quarkonium production
was the realization that fragmentation dominates at sufficiently large
$P_T$ \cite{fragm} which indicates that most charmonium at large $P_T$
is produced by the fragmentation of individual large $P_T$ partons.
The fragmentation functions are calculated in  CSM.
Including this fragmentation mechanism indeed brings the theoretical
predictions for prompt $\jpsi$ production  at the Tevatron to within a
factor of 3 of the data \cite{bdfm}. But the prediction for
the $\psi^{\prime}$ production cross section remains a factor
of 30 below the data even after including the fragmentation contribution
(the $\psi^{\prime }$``surplus'' problem).
Furthermore, the presence of the logarithmic infrared divergences in
the production cross sctions for P-wave charmonium states and the
annihilation rate for $\chi_{cJ} \to q {\bar q} g $ indicate CSM
is incomplete. All these indicate that important production
mechanism beyond CSM needs to be included \cite{pwcom}\cite{gcofrag}.
So the color-octet mechanism \cite{review} is proposed which is based on the
factorization formalism of NRQCD \cite{comfact}\cite{nrqcd}.
Contrary to the basic assumption of CSM, the heavy quark pair in a
color-octet state can make a transition into physical quarkonium state
through soft color interactions.

\par
Although the color-octet fragmentation picture of heavy quarkonium production
\cite{gcofrag} has provided valuable insight, the approximation that enter
into fragmentation computations break down when a quarkonium's energy
becomes comparable to its mass. The fragmentation predictions for heavy
quarkonium production are therefore unrealiable at low $P_T$.
Based upon the above several recently developed ideas in
heavy quarkonium physics, Cho and Leibovich\cite{cho} identify a large class
of color-octet diagrams that mediate quarkonia production at all energies,
which reduce to the dominant set of gluon fragmentation graphs in the high
$P_T$ limit. By fitting the data on heavy quarkonia production at the
Tevatron, numerical values for the long distance matrix
elements are extracted\cite{cho,beneke}, which are generally consistent with
NRQCD power  scaling rules \cite{power}.

\par
In order to convincingly establish the color-octet machanism, it is important
to test whether the same matrix elements be able to explain heavy quarkonium
production in other high energy processes, such
as inclusive $\jpsi$ production in $e^+ e^-$ annihilation  and
inelastic $\jpsi $ photoproduction at HERA {\it et al. }.
Other machanism for heavy quarkonium production can be found in
\cite{cem,sci}.

\subsection{The hard subprocesses for $\jpsi + jet$ production}

$\jpsi $ is described within the NRQCD framework in terms of Fock state
decompositions as
\ba
|\jpsi \rangle &=& O(1)~ |c{\bar c}[^3S_{1}^{(1)}] \rangle +
            O(v) |c{\bar c}[^3P_{J}^{(8)}] g \rangle  \nonumber \\
        & & + O(v^2) |c{\bar c}[^1S_{0}^{(8)}] g \rangle +
            O(v^2) |c{\bar c}[^3S_{1}^{(1,8)}] g g \rangle  \nonumber \\
        & & + O(v^2) |c{\bar c}[^3P_{J}^{(1,8)}] g g\rangle + \cdots ,
\ea
where the $c{\bar c}$ pairs are indicated within the square brackets in
spectroscopic notation. The pairs' color states are indicate by singlet (1)
or octet (8) superscripts. The color octet $c{\bar c}$ state can
make a transition into a physical $\jpsi$ state by soft chromoelectric
dipole (E1) transition(s) or chromomagnitic dipole M1 transition(s)
\beq
(c {\bar c})[^{2S + 1}L_{j}^{(8)}] \to \jpsi  .
\eeq
NRQCD factorization scheme \cite{comfact} has been established to
systematically separate high and low energy scale interactions.
It is based upon a bouble power series expansion in the strong interaction
fine structure constant $\alpha _s $ and the small
velocity parameter $v$. The production of a
$ (c {\bar c})[^{2S + 1}L_{j}^{(1,8)}] $ pair with separation less than
or of order $\frac{1}{m_c}$ can be calculated perturbatively.
The long distance
effects for the produced almost point-like $c{\bar c}$ to form the bound
state are isolated into nonperturbative matrix elements. Furthermore,
 NRQCD power counting rules can be exploited to determine the dominant
contributions to various quarkonium processes\cite{power}.
For direct $\jpsi $ production, the color-octet matrix elements,
$\mtso , \moso$ and $ \mtpj / m_c^2 $ are all scaling as $m_c^3 v_c^7 $.
So these color-octet contributions to $\jpsi $ production must be included
for consistency.

\par
On the partonic level, $\jpsi + jet$
production are composed of the photon gluon fusion, which are
sketched in Fig.2.  These are
\ba
\gamma + g & \to & g + (c{\bar c})[^3S_{1}^{(1)}, ^3S_{1}^{(8)}]~, 
~{\rm Fig.2.(a)-(f)};  \nonumber \\
\gamma + g & \to & g + (c{\bar c})[^1S_{0}^{(8)}, ^3P_{J}^{(8)}]~, 
~{\rm Fig.2.(a)-(h)}. \nonumber
\ea
The quark initiated subprocesses($\gamma q({\bar q})$ channel) are strongly
suppressed and will be neglected further.

\par
The color SU(3) coefficients are given by
\beq
\langle 3i;{\bar 3}j | 1 \rangle =\delta_{ji}/\sqrt{3},~
\langle 3i;{\bar 3}j | 8a \rangle =\sqrt{2} T^{a}_{ji}.
\eeq
Some identities involving the traces of the color matrices are useful
when the matrix elements are squared and spin, color averaged:
\ba
\frac{1}{8} {\rm Tr}(T^b T^c) {\rm Tr}^* (T^b T^c) \frac{1}{\sqrt{3}}
\frac{1}{\sqrt{3}}
&=& \frac{1}{12}, \nonumber \\
\frac{1}{8} {\rm F_{c}^{(a)-(c)}} {\rm F_{c}^{* (a)-(c)}} &=& \frac{7}{12},
\nonumber \\
\frac{1}{8} {\rm F_{c}^{(a)-(c)}} {\rm F_{c}^{* (d)-(f)}} &=& \frac{-1}{6},
\nonumber \\
\frac{1}{8} {\rm F_{c}^{(d)-(f)}} {\rm F_{c}^{* (a)-(c)}} &=& \frac{-1}{6},
\nonumber \\
\frac{1}{8} {\rm F_{c}^{(d)-(f)}} {\rm F_{c}^{* (d)-(f)}} &=& \frac{7}{12},
\nonumber \\
\frac{1}{8} {\rm F_{c}^{(g)-(h)}} {\rm F_{c}^{* (h)-(h)}} &=& \frac{3}{2},
\nonumber \\
\frac{1}{8} {\rm F_{c}^{(g)-(h)}} {\rm F_{c}^{* (a)-(c)}} &=& \frac{- 3 i }{4},
\nonumber \\
\frac{1}{8} {\rm F_{c}^{(g)-(h)}} {\rm F_{c}^{* (d)-(f)}} &=& \frac{3 i }{4},
\nonumber \\
\frac{1}{8} {\rm F_{c}^{* (g)-(h)}} {\rm F_{c}^{ (a)-(c)}} &=& \frac{3 i }{4},
\nonumber \\
\frac{1}{8} {\rm F_{c}^{* (g)-(h)}} {\rm F_{c}^{ (d)-(f)}} &=& \frac{- 3 i }{4}.
\ea
\noindent
Where
\ba
{\rm F_{c}^{(a)-(c)}}&=& \sqrt{2} {\rm Tr}(T^a T^b T^c), \nonumber \\
{\rm F_{c}^{(d)-(f)}}&=& \sqrt{2} {\rm Tr}(T^a T^c T^b), \nonumber \\
{\rm F_{c}^{(g)-(h)}}&=& \sqrt{2} {\rm Tr}(T^a T^d) f_{dbc} \nonumber \\
&=&\frac{\sqrt{2}}{2} f_{abc}.
\ea

\par
Note that the hard subprocesses that producing color-octet
$(c{\bar c})[^1S_{0}^{(8)}, ^3P_{J}^{(8)}]$ pair involve triple-gluon
coulping (Two of them are external gluons). This situation implies that
we cannot use the trick of replacing the sum over gluon polarization states
by 
\beq
\nonumber
\sum_{\lambda } \varepsilon _{\alpha } (\lambda)
\varepsilon _{\alpha^{\prime}}^{*} (\lambda)
\to - g_{\alpha \alpha^{\prime}} ,
\eeq

\noindent
when we use the Feynman gauge for the gluon propagator\cite{field}.
One way to obtain the correct result is insist that the polarization
states of the two external gluons are physical ({\it i.e.,} transverse).
This is accomplished by the projection
\ba
~&&\sum_{\lambda } \varepsilon _{\alpha } (\lambda)
\varepsilon _{\alpha^{\prime}}^{*} (\lambda)  \nonumber \\    
~&&= -
\Bigg[ g_{\alpha \alpha^{\prime}} - \frac{n_{\alpha}k_{\alpha^{\prime}} +
n_{\alpha^{\prime}}k_{\alpha}} {(n\dot k)} + \frac{n^2 k_{\alpha}
k_{\alpha^{\prime}}} {(n\dot k)^2}\Bigg],
\ea

\noindent
where $n$ is an arbitrary 4-vector and $k$ is the gluon 4-momentum.
This is analogous to using an axial ({\it i.e.}physical) gauge for
the gluon propagator.
Our choice for the 4-vector $n$ is $n = g_2 $ for the incoming gluon and
$ n = g_1 $ for the outgoing gluon, $ g_1, g_2 $ are the 4-momenta
of the incoming and outgoing gluons respectively.
 
\par
With all ingredients set as above, we obtain the
spin, color average-squared matrix elements
$ \overline{\Sigma }|M(\gamma + g \to 
(c {\bar c})[^{2S + 1}L_{j}^{(1; 8)}] + g)|^2 $.  Our analytic
expressions are consistent  with  \cite{berg} and the results
calculated from Eqs. (11),(15) and appendix B of \cite{bkv}, but different
from the expressions of \cite{kim}. For example,   
the expression of $ \overline{\Sigma }|M(\gamma + g \to 
(c {\bar c})[^1 S_{0}^{(8)}] + g)|^2 $ of\cite{kim} is different from our
analytic expression by a term

\beq
\frac{6 \hat{s} \hat{u} ( e e_c g_s^2)^2 }
{\hat{t} (\hat{s}+\hat{u})^2 },
\eeq

\noindent
which is exactly the contribution of
photoh-ghost scattering diagram ({\it i.e.} the contribution of the
unphysical polarization states of the external gluons) and should be
subtracted from Eq. (A1) of \cite{kim}. Here
\beq
\hat{s} = (q + g_1)^2 ,~\hat{t} = (q - P)^2 ,~\hat{u} = (g_1 - P)^2 .
\eeq
\noindent
In appendix, we give explicit expressions of
$ \overline{\Sigma }|M(\gamma + g \to 
(c {\bar c})[^{2S + 1}L_{j}^{(1; 8)}] + g)|^2 $ for $\jpsi$ production. 

\par
Known the spin, color average-squared matrix elements of the
hard subprocesses, in the NRQCD framework, we obtain the
subcross sections. The color-singlet photon-gluon fusion
contribution to $\jpsi + jet $ production is well  known \cite{berg}:
\ba
\label{sg3s11}
~&&\frac{d\hat{\sigma}} {d\hat{t}}[\gamma + g  \to
   (c{\bar c})[^3S_{1}^{(1)}] + g \to \jpsi + jet]  \nonumber \\
~&&~~~~= \frac{1}{16 \pi \hat{s}^2 } 
\overline{\Sigma }
    |M(\gamma + g \to  (c{\bar c})[^3S_{1}^{(1)}] + g)|^2 \nonumber \\
~&&~~~~~    \times \frac{1}{18 m_c} \mtss .
\ea

\noindent
where $\mtss $ is the color-singlet matrix element which is related to the
lepton decay width of $\jpsi$.  The average-squared matrix element
$ \overline{\Sigma }|M(\gamma + g \to 
(c {\bar c})[^3 S_{1}^{(8)}] + g)|^2 $  can be obtained from
$ \overline{\Sigma }|M(\gamma + g \to (c {\bar c})[^3 S_{1}^{(1)}] + g)|^2 $
by taking into account of different color factor.
The color-octet $ (c {\bar c})[^3 S_{1}^{(8)}] $ contribution is

\ba
\label{sg3s18}
~&&\frac{d\hat{\sigma}} {d\hat{t}}[\gamma + g  \to
    (c{\bar c})[^3S_{1}^{(8)}] + g  \to \jpsi + g]  \nonumber \\
~&&~~~~= \frac{1}{16 \pi \hat{s}^2 } \frac{15}{6}
\overline{\Sigma }
    |M(\gamma + g \to (c{\bar c})[^3S_{1}^{(1)}] + g)|^2 \nonumber \\
~&&~~~~~    \times \frac{1}{24 m_c} \mtso .
\ea

\noindent
The color-octet $ (c {\bar c})[^3 S_{0}^{(8)}] $ and
$ (c {\bar c})[^3 P_{J}^{(8)}] $ contributions are 
\ba
\label{sgsp08}
~&&\frac{d\hat{\sigma}} {d\hat{t}}[\gamma + g  \to
   (c{\bar c})[^1S_{0}^{(8)}] + g \to \jpsi + jet] \nonumber \\
~&&~~~~= \frac{1}{16 \pi \hat{s}^2 } \overline{\Sigma }
    |M(\gamma + g \to (c{\bar c})[^1S_{0}^{(8)}] + g)|^2  \nonumber \\
~&&~~~~~    \times \frac{1}{8 m_c} \moso , \nonumber \\ 
~&&\frac{d\hat{\sigma}} {d\hat{t}}[\gamma + g  \to
   (c{\bar c})[^3P_{J}^{(8)}] + g \to \jpsi + jet] \nonumber \\
~&&~~~~= \frac{1}{16 \pi \hat{s}^2 } \sum_{J} \overline{\Sigma }
    |M(\gamma + g \to (c{\bar c})[^3P_{J}^{(8)}] + g)|^2  \nonumber \\
~&&~~~~~    \times \frac{1}{8 m_c} \mtpo ,
\ea
where the heavy quark spin symmetry
\beq
\mtpj = (2 J + 1 ) \mtpo
\eeq
is exploited.

\subsection{The $z$ and $ P_T $ distributions of $\jpsi $ }

Now we consider the $z$ and $P_T $ distribution of $\jpsi $ produced in
process Eq.(\ref{proc}).
Based on the factorization theorem for the lepton induced hard diffractive
hard scattering\cite{jccollins}, the differential cross section can be
expressed in terms of a diffractive parton
distribution\cite{kunszt,berera} as

\ba
\label{dsigma}
d\sigma &=& \frac{d f_{g/p}^{diff} ( x_1, x_{{\rm I\!P}}, t, \mu) }
            {dx_{{\rm I\!P}} dt}  dx_{{\rm I\!P}} dt \cdot \nonumber \\
~&~& \frac{d \hat{\sigma} (\gamma +g \to \jpsi + jet)} {d\hat{t}}
dx_1 d \hat{t}.
\ea

\noindent
Here

\beq
\frac{d f_{g/p}^{diff} ( x_1, x_{{\rm I\!P}}, t, \mu) }
     {dx_{{\rm I\!P}} dt}  dx_1
\eeq

\noindent
represents the probability of finding in the proton a gluon carrying
momentum faction
$ x_1$, while leaving  the proton intact except for been diffractively
scattered with the momentum transfer  $(x_{{\rm I\!P}}, t)$.
$x_{{\rm I\!P}}$ is the fractional momentum loss of the diffracted proton,
{\it i.e.,} $x_{{\rm I\!P}} \simeq
(p_{1 z} - p_{1 z}^{\prime})/p_z$,
and $t$ is the invariant momentum transfer for the diffracted proton,
{\it i.e.,} $t = (p_1 - p_{1}^{\prime})^2$.

\par
We now consider the kinematics. It is convenient to introduce the variable

\beq
z \equiv \frac{p_1 \cdot P}{p_1 \cdot q}.
\eeq

\noindent
In the proton rest frame $ z = E_{\psi}/E_{\gamma}$. In terms of $z$ and
$P_T$ (the transverse momentum of $\jpsi$), the Mandelstam variables and
$x_1$ can be expressed as the following:
\ba
\label{hatstu}
\hat{s} &=& \frac{P_T^2} {z (1 - z)} +\frac{m_{\psi}^2} {z}, \nonumber \\
\hat{t} &=& - \frac{(1 - z) m_{\psi}^2 + P_T^2} {z}, \nonumber \\
\hat{u} &=& - \frac{P_T^2}{1 - z}, \nonumber \\
x_1     &=& \frac{\hat{s}} {s_{\gamma p}}.
\ea

\noindent
Here, $ s_{\gamma p} = (q+p)^2 $. The double differential cross section is

\ba
\frac{d\sigma}{dz dP_T} &=& \int_{x_1}^{x_{\rm I\!P max}} dx_{\rm I\!P}
                            \int_{-1}^{0} dt 
               \frac{d f_{g/p}^{diff} ( x_1, x_{{\rm I\!P}}, t, \mu) }
            {dx_{{\rm I\!P}} dt} \cdot \nonumber \\
~&~& \frac{d \hat{\sigma} (\gamma +g \to \jpsi + jet)} {d\hat{t}}
     J(\frac{x_1 \hat{t} }{z P_T} ),
\ea

\noindent
where the Jacobian can obtain from Eq.(\ref{hatstu}),
\beq
J(\frac{x_1 \hat{t} }{z P_T} )
= \frac { 2 P_T \hat{s} }  { z(1-z) s_{\gamma p}}.
\eeq

\noindent
The allowed regions of $ z, P_T $ are given by
\ba
0           &\leq P_T \leq & \sqrt{ (1-z) (z x_{\rm I\!P  max} s_{\gamma p}
                                    - m_{\psi}^{2})}, \nonumber \\
z_{\rm min} &\leq  z  \leq & z_{\rm max},
\ea
\noindent
with
\ba
z_{\rm max}&=& \frac{1}{2 x_{\rm I\!P  max} s_{\gamma p} }
               \Bigg[ x_{\rm I\!P  max} s_{\gamma p} + m_{\psi}^{2} \nonumber \\
~          &~&  + \sqrt{ (x_{\rm I\!P  max} s_{\gamma p} - m_{\psi}^{2})^2
                        - 4 x_{\rm I\!P  max} s_{\gamma p} P_T^2} \Bigg],
                        \nonumber \\
z_{\rm min}&=& \frac{1}{2 x_{\rm I\!P  max} s_{\gamma p} }
             \Bigg[x_{\rm I\!P  max} s_{\gamma p} + m_{\psi}^{2} \nonumber \\
~          &~&  - \sqrt{ (x_{\rm I\!P  max} s_{\gamma p} - m_{\psi}^{2})^2
                        - 4 x_{\rm I\!P  max} s_{\gamma p} P_T^2} \Bigg].
\ea

\noindent
In order to suppress the Reggon contributions, we set
$x_{\rm I\!P  max} = 0.05$ as usual.

\section{Numerical results and discussions}
For numberical predictions, we use
$m_c = 1.5 GeV, \Lambda_4 = 235 MeV $, and set the factoriztion
scale and the renormalization scale both equal to the transverse mass
of $\jpsi$, {\it i.e.}, $\mu^2 = m_T^2 = (m_{\psi}^2 + P_T^2)$ .
For the color-octet matrix elements $\mtso , \moso$ and $\mtpo $ we
use the values determined by Beneke and Kr$\ddot{a}$mer \cite{beneke}
from fitting the direct $\jpsi$ production data at
the Tevatron using GRV LO parton  distribution functions\cite{grv},

\ba
\label{matrix}
&& \mtso = 1.12 \times 10^{-2} GeV^3,  \nonumber \\
&& \moso + \frac{3.5}{m_c^2} \mtpo  \nonumber \\
&& = 3.90 \times 10^{-2} GeV^3 .
\ea
Since the matrix elements $\moso $ and $\mtpo$ are not determined separately,
we choose
\ba
\moso       &=& 1.0 \times 10^{-2} {\rm GeV}^3, \nonumber \\
\mtpo/m_c^2 &=& 8.3 \times 10^{-3} {\rm Gev}^3,
\ea

\noindent
allowed by Eq.(\ref{matrix}), and in accordance with NRQCD power counting
rule.  The value of the color-singlet matrix element is taken to be
$\mtss = 1.16 {\rm GeV}^3$.  The diffractive gluon distribution function

\beq
\nonumber
\frac{d f_{g/p}^{diff} ( x_1, x_{{\rm I\!P}}, t, \mu) }
{dx_{{\rm I\!P}} dt}
\eeq

\noindent
at $\mu^2 = 4 {\rm GeV}^2$ has been extracted from data on DDIS and
on diffractive photoproduction of jets at HERA in \cite{alvero}, 
\ba
&& \frac{d f_{g/p}^{diff} ( x_1, x_{{\rm I\!P}}, t, \mu^2 = 4 {\rm GeV}^2) }
   {dx_{{\rm I\!P}} dt}   \nonumber \\
&& = \frac{9 \beta_0^2}{4 \pi^2} \Bigg[
     \frac{4 m_p^2 - 2.8 t}{4 m_p^2 -t} \bigg( \frac{1}{1 - t/0.7}\bigg)^2
     \Bigg] x_{\rm I\!P}^{- 2 \alpha (t)} \cdot \nonumber \\
&& ~~~~a_g (1 - x_1/x_{\rm I\!P}), 
\ea

\noindent
and is evolved in $\mu^2$ according to the DGLAP evolution
equations\cite{jccollins}.
Here $ \beta_0 = 1.8 {\rm GeV}^{-1}$, $\alpha (t) = 1.14 + 0.25 t$,
$ a_g = 4.5 \pm 0.5$.
We use the central value of $a_g$ for numberical calculation.
In order to suppress the elastic $\jpsi$ photoproduction
(with or without proton dissociation) contribution and higher-twist
corrections, a $P_T$ cut $P_T \geq 1.0 {\rm GeV}$ is imposed.

\par
In Fig.3 we show the $z$ distribution $ \frac{d \sigma} {d z} $ of
$\jpsi$ produced in process Eq.(\ref{proc}) at HERA at a typical
energy $\sqrt{s_{\gamma p}} = 100 {\rm GeV}$ and with a $P_T$ cut
$P_T \geq 1.0 {\rm GeV}$.
The thin dash-dotted and short dashed lines
represent the color-singlet cotribution and the sum of the color-singlet
and color-octet contributions respectivly.  We observe that for $z > 0.3$
the color-singlet contribution is about order of $ 10^0 {\rm nb}$, so this
process can be studied at HERA with present integrated luminosity.
For completeness, in Fig.3, we also show the $z$ distribution of
$\jpsi$ inelastic production $ \gamma + p \to \jpsi + jet + X $
in the same kinematic region (upper part), in this case,
the thick solid and dotted lines curves are the sum of the color-singlet
and color-octet contributions and the color-singlet contribution
respectively. In both case, including the color-octet contribution,
the $z$ distribution is strongly enhanced in high $z$ region due to
the gluon propagator in the color-octet channel (Fig.2.(g)-(h)).
In inelastic case, this behavior of rapid growing at high $z$ does not
agree with the HERA data\cite{h1-jpsi,zeus-jpsi}.

\par
In Fig.4, we show the $P_T$ distribution $ \frac{d \sigma} {d P_T} $
at HERA with $\sqrt{s_{\gamma p}} = 100 {\rm GeV}$ and a $z$ cut
$z\leq 0.8$. The code for the curves are the same as Fig.3.
We find that for $\jpsi+jet$ diffractive photoproduction in process
Eq.(\ref{proc}) the color-octet contributions are larger than the
color-singlet contribution for $P_T > 2.6 {\rm GeV}$.

\par
In Fig.5, we show the $\jpsi+jet$ diffractive photoproduction cross section
as a function of $\sqrt{s_{\gamma p}}$ with the cut, $P_T \geq 1.0 {\rm GeV},
z \leq 0.8 $. We observe that in the energy region considered the
color-singlet contribution is about order of $ 10^0 {\rm nb}$, so this
process can be studied at HERA with present integrated luminosity, and
can give valuable insights in the color-octet mechanism for heavy quarkonia
production.

\par
Recently, it was pointed out that the cross section with an additional
$ z $ cut, say, $z < 0.8$, {\it cannot} be reliably predicted in NRQCD
\cite{bkv,brw}, because the NRQCD expansion is singular at $z = 1$, only
an average cross section over a sufficiently large region close to $z = 1$
can be predicted. The $z$ distribution itself requires additional
non-perturbative information in the form of so-called shape functions.
These shape functions are also required to predict the $P_T$ distributions
with an additional upper $z$ cut, say, $ z < 0.8$, but not if $z$ is
integrated up to its kinematic maximum.
So in Fig.6, we show the the $P_T$ distribution $ \frac{d \sigma} {d P_T} $
at HERA with $\sqrt{s_{\gamma p}} = 100 {\rm GeV}$ and
with $z$ is
integrated from  its lower cut $0.2$ to its kinematic maximum.
The code for the curves are the same as Fig.4.
We find that for $\jpsi+jet$ diffractive photoproduction in process
Eq.(\ref{proc}), without the upper $z$ cut, the color-octet contributions
are dominated in the whole $P_T$ region considered, which exceed the
color-singlet contribution by almost an order of magnitude.
In Fig.7, the $\jpsi + jet$ diffractive photoproduction cross section
as a function of $\sqrt{s_{\gamma p}}$ with $z$ is integrated from its
lower cut $0.2$ to its kinematic maximum is shown.
We observe that in the energy region considered the
color-singlet contribution is in the region $ 1.4 {\rm nb} <
\sigma^{diff.} (singlet) < 2.4 {\rm nb}$, while including the color-octet
contributions, the cross section increased by about an order of magnitude.

\par
Experimentally, the nondiffractive and elastic (with or without proton
dissociation) background to the diffractive $\jpsi$ photoproduction we
studied must be dropped out in order to obtain useful information about
the heavy qarkonia production mechanism, this can be attained by performing
the rapid gap analysis together with a large $P_T$ cut, say,
$P_T \geq 1.0 {\rm GeV}$. If statisitics is not a limitation, it might
be preferable to use cut $P_T \geq 2 {\rm GeV}$.

\par
In conclusion, in this paper,
We present a study of $\jpsi + jet$ diffractive production in the direct
photon process at  HERA based on the factorization theorem for lepton-induced
hard diffractive scattering and the factorization formalism of
the nonrelativistic QCD (NRQCD) for quarkonia production.
Using the diffractive gluon distribution function
extracted from HERA data on diffractive deep inelastic scattering and
diffractive dijet photon production,
we show that this process  can be
studied at HERA with present integrated luminosity, and can give valuable
insights in the color-octet mechanism for heavy quarkonia production.

\vskip 1cm
\begin{center}
{\bf\large Acknowledgments}
\end{center}
This work is supported in part by the National Natural Science Foundation of
China, Doctoral Program Foundation of Institution of Higher Education of
China and Hebei Province Natural Science Foundation, China.

\newpage
\appendix
\centerline{APPENDIX: THE SPIN, COLOR AVERAGE-SQUARED MATRIX ELEMENTS }

\par
In this appendix, we give explicit expressions for the 
spin, color average-squared matrix elements of photon-gluon fusion
subprocesses. The results are obtained using the symbolic manipulations with
the aid of REDUCE package.

\ba
&&\overline{\Sigma }|M(\gamma + g \to  (c{\bar c})[^3S_{1}^{(1)}] + g)|^2
   \nonumber \\
&&= \frac{64}{12} (e e_c g_s^2)^2 m_{\psi}^2 
    \frac{\hat{s}^2 (\hat{t}+\hat{u})^2 + \hat{t}^2 (\hat{s}+\hat{u})^2 + \hat{u}^2 (\hat{s}+\hat{t})^2}
       { (\hat{t}+\hat{u})^2 (\hat{s} +\hat{u})^2 (\hat{u} +\hat{t})^2 } \\
&& ~~~~~~~~~~~~~~~~~~~~~~~~~~~~~~~ \nonumber \\
&& ~~~~~~~~~~~~~~~~~~~~~~~~~~~~~~~~~~~~~~~~~~~~ \nonumber \\
&&\overline{\Sigma }
    |M(\gamma + g \to (c{\bar c})[^3S_{1}^{(8)}] + g)|^2
    \nonumber \\
&&= \frac{15}{6}
    \overline{\Sigma }
    |M(\gamma + g \to (c{\bar c})[^3S_{1}^{(1)}] + g)|^2   \\
&& ~~~~~~~~~~~~~~~~~~~~~~~~~~~~~~~ \nonumber \\
&& ~~~~~~~~~~~~~~~~~~~~~~~~~~~~~~~~~~~~~~ \nonumber \\
&& \overline{\Sigma }
    |M(\gamma + g \to (c{\bar c})[^1S_{0}^{(8)}] + g)|^2
    \nonumber \\
&& = \frac{24 \hat{s} \hat{u} (e e_c  g_s^2)^2 }
          {\hat{t} (\hat{s} + \hat{t})^2 (\hat{s} +
          \hat{u})^2 (\hat{t} + \hat{u})^2 }  \times \nonumber \\
&& ~~(\hat{s}^4 + 2 \hat{s}^3 \hat{t} + 2 \hat{s}^3 \hat{u} +
3 \hat{s}^2 \hat{t}^2 + 6 \hat{s}^2 \hat{t} \hat{u} + 3 \hat{s}^2 \hat{u}^2 +
2 \hat{s} \hat{t}^3  \nonumber \\
&& ~~+ 6 \hat{s} \hat{t}^2 \hat{u} + 6 \hat{s} \hat{t} \hat{u}^2 +
2 \hat{s} \hat{u}^3 + \hat{t}^4 + 2 \hat{t}^3 \hat{u} +
3 \hat{t}^2 \hat{u}^2 + 2 \hat{t} \hat{u}^3 + \hat{u}^4)  \nonumber \\
&& = \frac{24 \hat{s} \hat{u} (e ec  g_s^2)^2 }
    {\hat{t} (\hat{s} + \hat{t})^2 (\hat{s} + \hat{u})^2 (\hat{t} +
    \hat{u})^2 }
     \{ [\hat{u}^2 + (\hat{s}+\hat{u}+\hat{t})(\hat{s}+\hat{t})]^2 -
     2 \hat{s} \hat{t} (\hat{s}+\hat{t})^2 +
           \hat{s}^2 \hat{t}^2\}  \\
&& ~~~~~~~~~~~~~~~~~~~~~~~~~~~~~~~ \nonumber \\
&& ~~~~~~~~~~~~~~~~~~~~~~~~~~~~~~~ \nonumber \\
&& \sum_{J}\overline{\Sigma}
    |M(\gamma + g \to (c{\bar c})[^3P_{J}^{(8)}] + g)|^2
    \nonumber \\
&& = \frac {96 (e e_c g_s^2)^2}
{\hat{t} m_{\psi}^{2} (\hat{s} + \hat{t})^3 (\hat{s} + \hat{u})^3 (\hat{t} +
\hat{u})^3 } \times \nonumber \\
&& ~~(7 \hat{s}^7 \hat{t} \hat{u} + 7 \hat{s}^7 \hat{u}^2 +
25 \hat{s}^6 \hat{t}^2 \hat{u} +  38 \hat{s}^6 \hat{t} \hat{u}^2 +
21 \hat{s}^6 \hat{u}^3 + 2 \hat{s}^5 \hat{t}^4 +
47 \hat{s}^5 \hat{t}^3 \hat{u} + 88 \hat{s}^5 \hat{t}^2 \hat{u}^2 +
78 \hat{s}^5 \hat{t} \hat{u}^3  \nonumber \\
&& ~~+ 35 \hat{s}^5 \hat{u}^4 + 4 \hat{s}^4 \hat{t}^5 +
63 \hat{s}^4 \hat{t}^4 \hat{u} + 132 \hat{s}^4 \hat{t}^3 \hat{u}^2 +
156 \hat{s}^4 \hat{t}^2 \hat{u}^3 + 98 \hat{s}^4 \hat{t} \hat{u}^4 +
35 \hat{s}^4 \hat{u}^5 + 2 \hat{s}^3 \hat{t}^6 +
47 \hat{s}^3 \hat{t}^5 \hat{u}   \nonumber \\
&& ~~+ 136 \hat{s}^3 \hat{t}^4 \hat{u}^2 + 190 \hat{s}^3 \hat{t}^3 \hat{u}^3
+156 \hat{s}^3 \hat{t}^2 \hat{u}^4 + 78 \hat{s}^3 \hat{t} \hat{u}^5 +
21 \hat{s}^3 \hat{u}^6 + 13 \hat{s}^2 \hat{t}^6 \hat{u} +
70 \hat{s}^2 \hat{t}^5 \hat{u}^2 + 136 \hat{s}^2 \hat{t}^4 \hat{u}^3
 \nonumber \\
&& ~~+ 132 \hat{s}^2 \hat{t}^3 \hat{u}^4 + 88 \hat{s}^2 \hat{t}^2 \hat{u}^5
+ 38 \hat{s}^2 \hat{t} \hat{u}^6 + 7 \hat{s}^2 \hat{u}^7 +
13 \hat{s} \hat{t}^6 \hat{u}^2 +  47 \hat{s} \hat{t}^5 \hat{u}^3 +
63 \hat{s} \hat{t}^4 \hat{u}^4 + 47 \hat{s} \hat{t}^3 \hat{u}^5 \nonumber \\
&& ~~+ 25 \hat{s} \hat{t}^2 \hat{u}^6 +
7 \hat{s} \hat{t} \hat{u}^7 + 2 \hat{t}^6 \hat{u}^3 +
4 \hat{t}^5 \hat{u}^4 + 2 \hat{t}^4 \hat{u}^5)
\ea

\newpage
\begin{figure}
\centerline{\epsfig{file=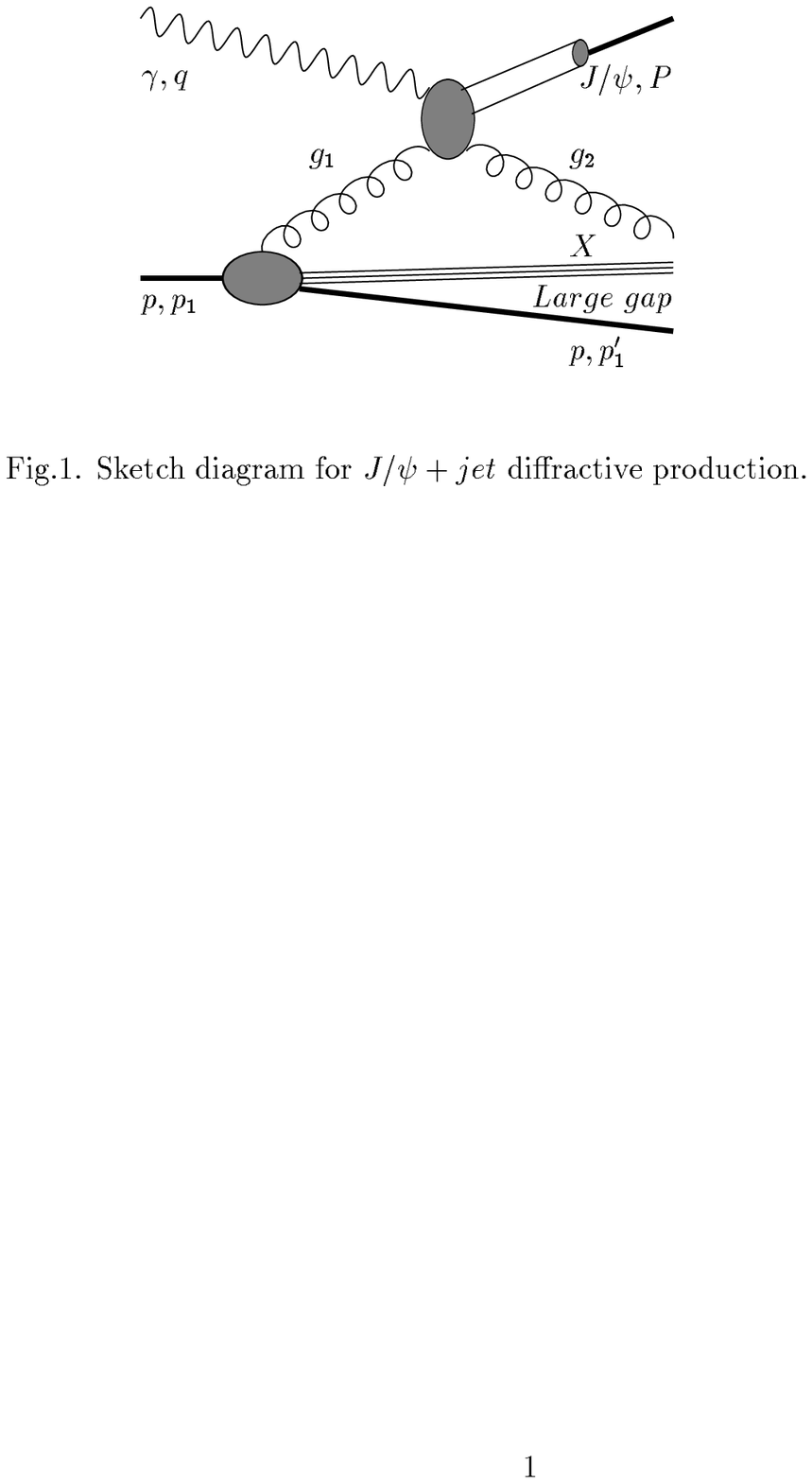,width=17cm}}

\centerline{\epsfig{file=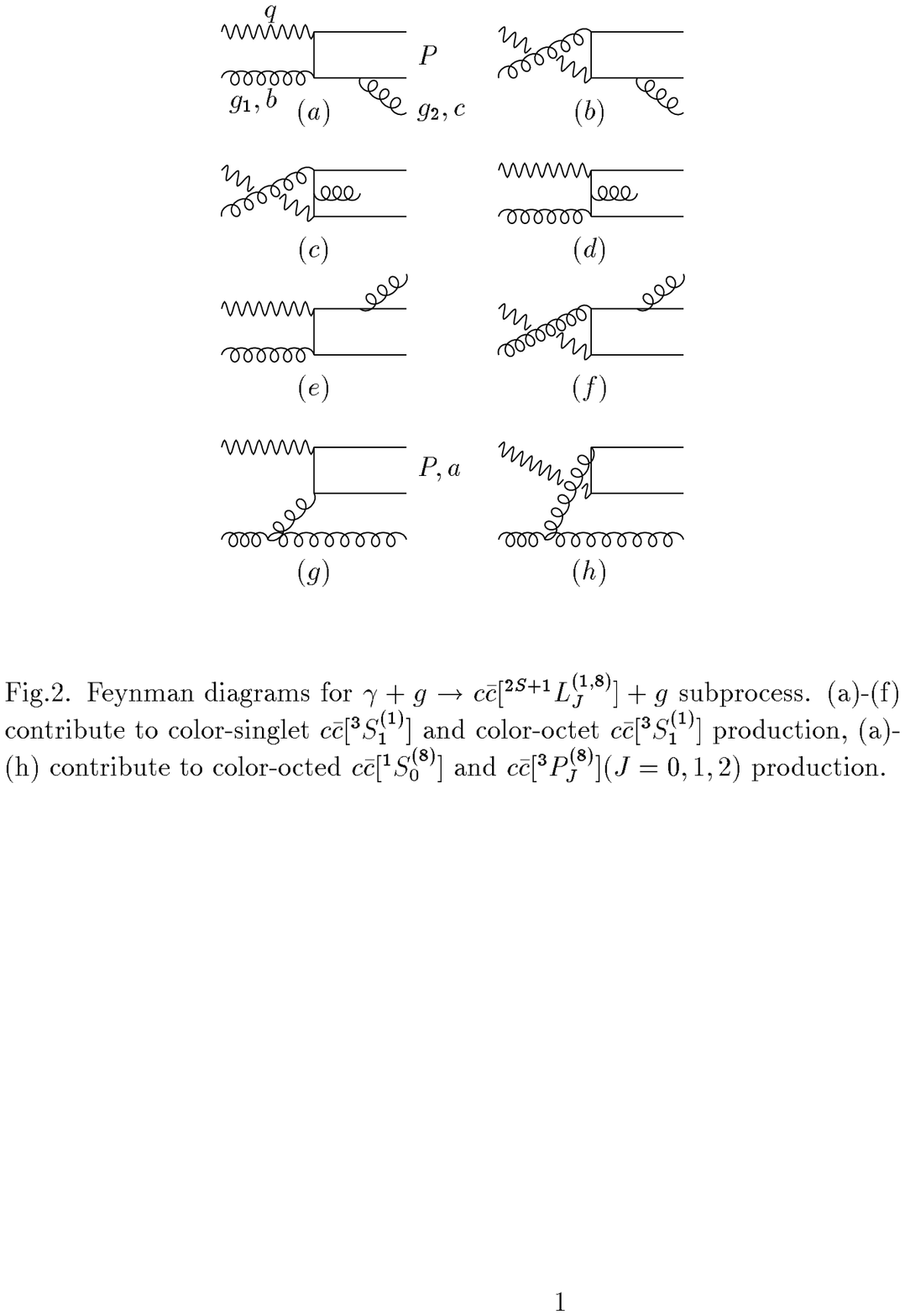,width=17cm}}

\centerline{\epsfig{figure=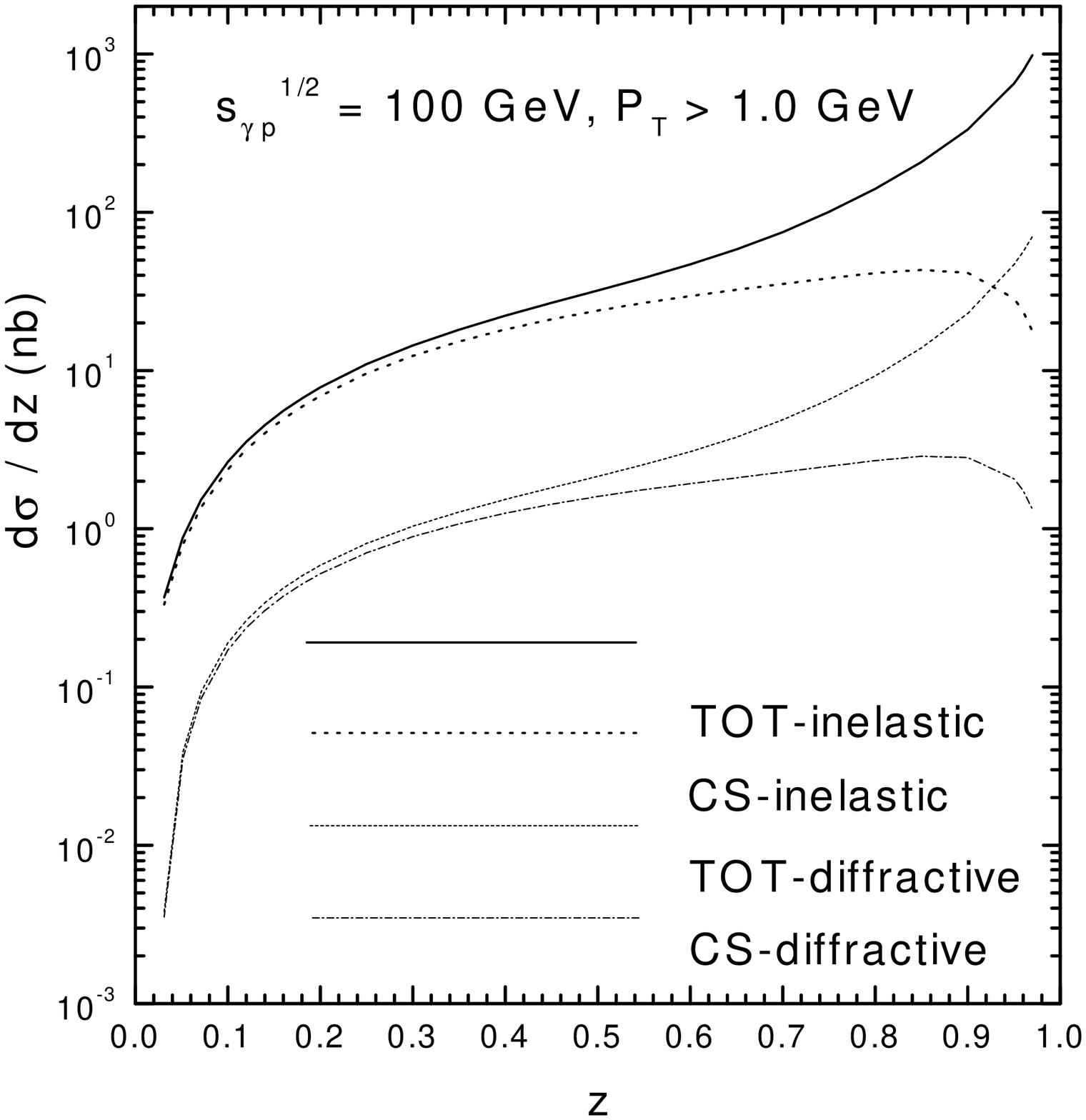,width=90mm}}
Fig.3. $z$ distribution $ d\sigma /dz$ for $\jpsi + jet$ production at
HERA at $\sqrt{s_{\gamma p}} = 100 {\rm GeV}$,
integrated over the $\jpsi$ transverse momentum $P_T$ with a lower
$P_T$ cut $P_T \geq 1.0 {\rm GeV}$.
The thin dash-dotted and short dashed lines
represent the color-singlet cotribution and the sum of the color-singlet
and color-octet contributions to
$\jpsi + jet$ diffractive photoproduction $\gamma + p \to \jpsi + jet + p
+ X$ respectivly.
The thick solid and dotted lines curves are the sum of the color-singlet
and color-octet contributions and the color-singlet contribution to
inelastic $\jpsi + jet$ photoproduction $\gamma + p \to \jpsi + jet + X$
respectively.

\newpage
\centerline{\epsfig{figure=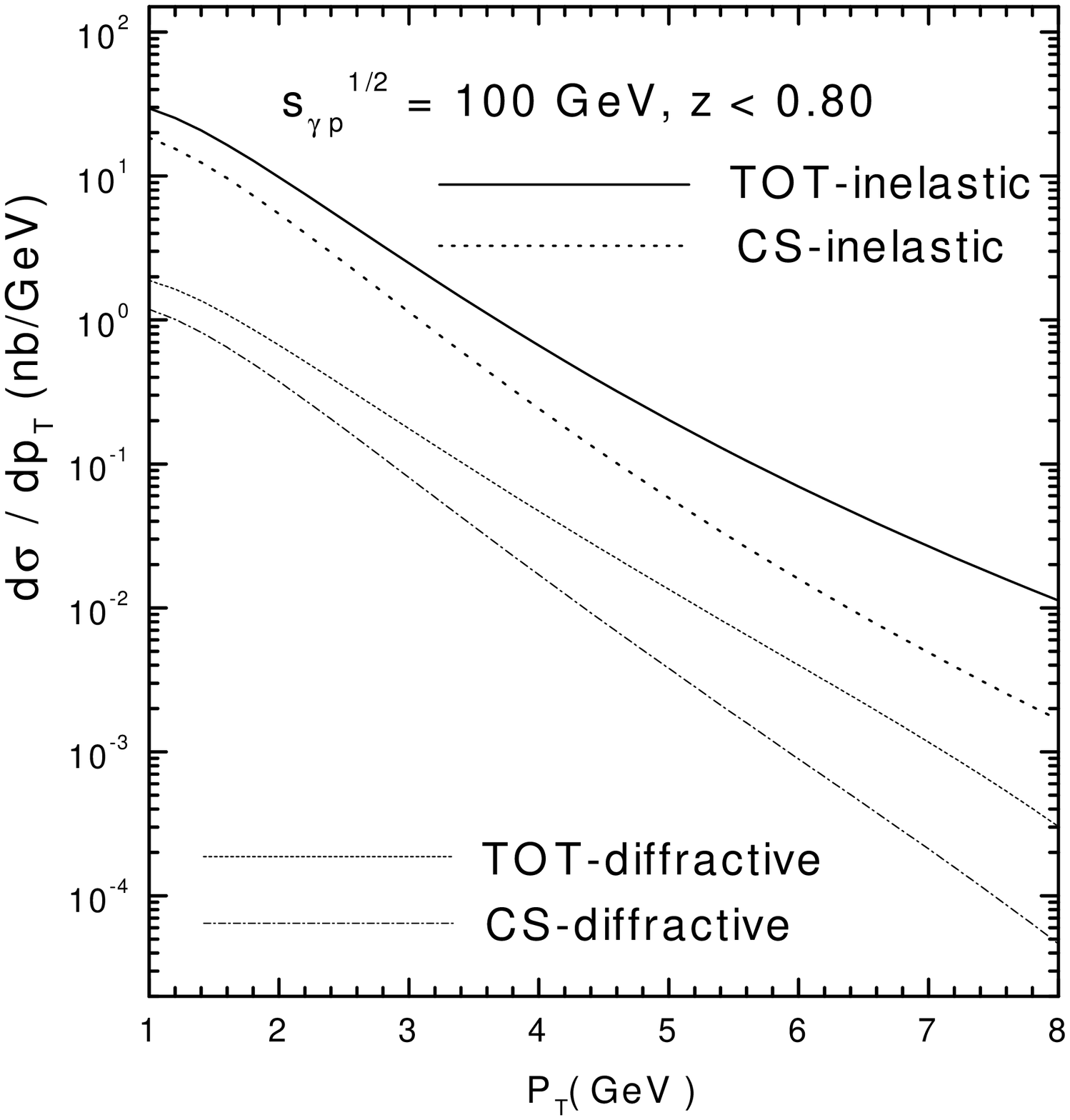,width=90mm}}
Fig.4. Transverse momentum ($P_T$) distribution $ d\sigma /dP_T $
for $\jpsi + jet$ production at HERA at
$\sqrt{s_{\gamma p}} = 100 {\rm GeV}$,
integrated over $z$ with a upper
$z$ cut $z \leq 0.8$.
The thin dash-dotted and short dashed lines
represent the color-singlet cotribution and the sum of the color-singlet
and color-octet contributions to
$\jpsi + jet$ diffractive photoproduction $\gamma + p \to \jpsi + jet + p
+ X$ respectivly.
The thick solid and dotted lines curves are the sum of the color-singlet
and color-octet contributions and the color-singlet contribution to
inelastic $\jpsi + jet$ photoproduction $\gamma + p \to \jpsi + jet + X$
respectively.

\newpage
\centerline{\epsfig{figure=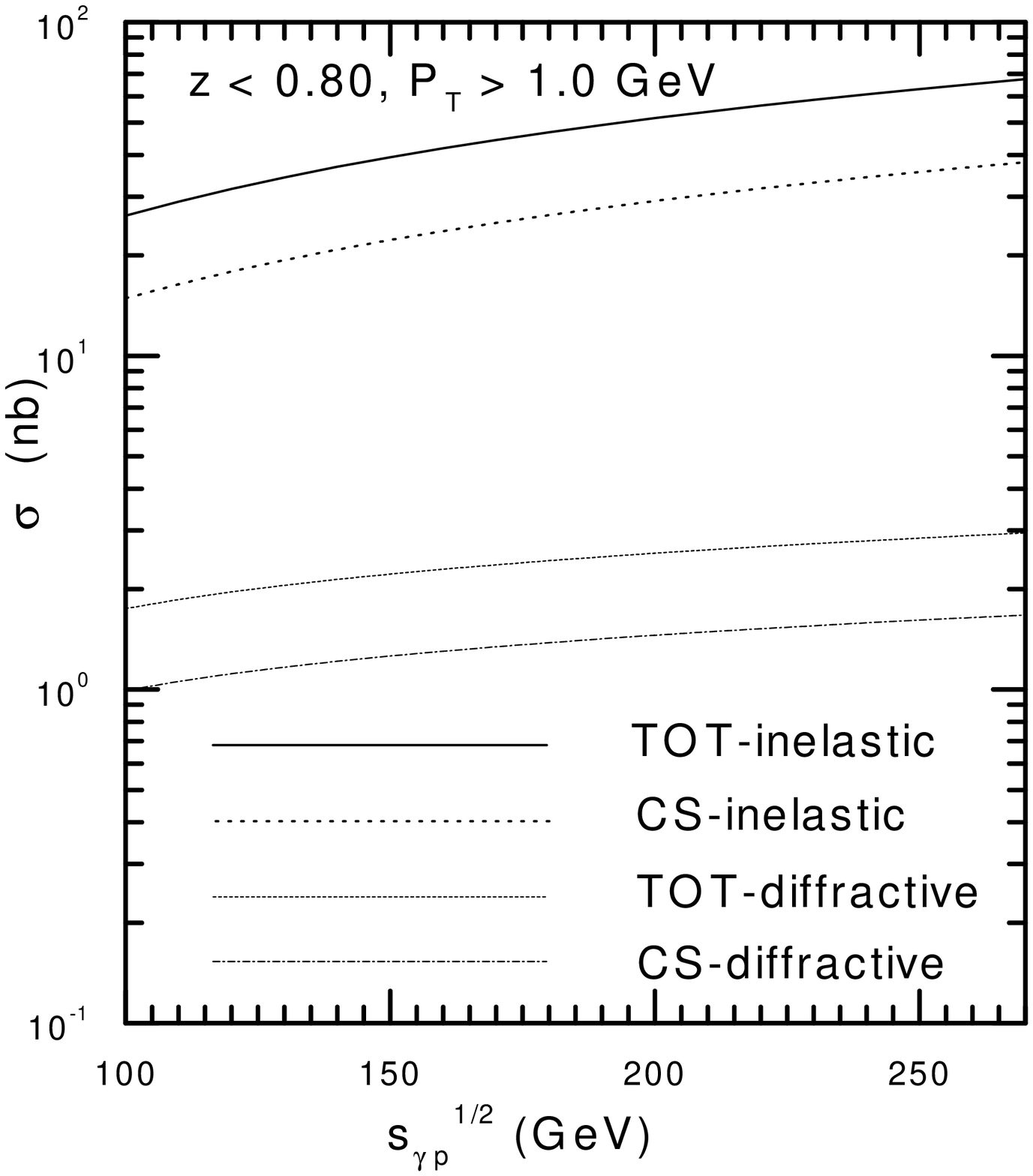,width=90mm}}
Fig.5. Total $\jpsi + jet$ photoproduction cross section for
$P_T \geq 1.0 {\rm GeV}, z\leq 0.8$ as a function of
$\sqrt{s_{\gamma p}}$.
The thin dash-dotted and short dashed lines
represent the color-singlet cotribution and the sum of the color-singlet
and color-octet contributions to
$\jpsi + jet$ diffractive photoproduction $\gamma + p \to \jpsi + jet + p
+ X$ respectivly.
The thick solid and dotted lines curves are the sum of the color-singlet
and color-octet contributions and the color-singlet contribution to
inelastic $\jpsi + jet$ photoproduction $\gamma + p \to \jpsi + jet + X$
respectively.

\newpage
\centerline{\epsfig{figure=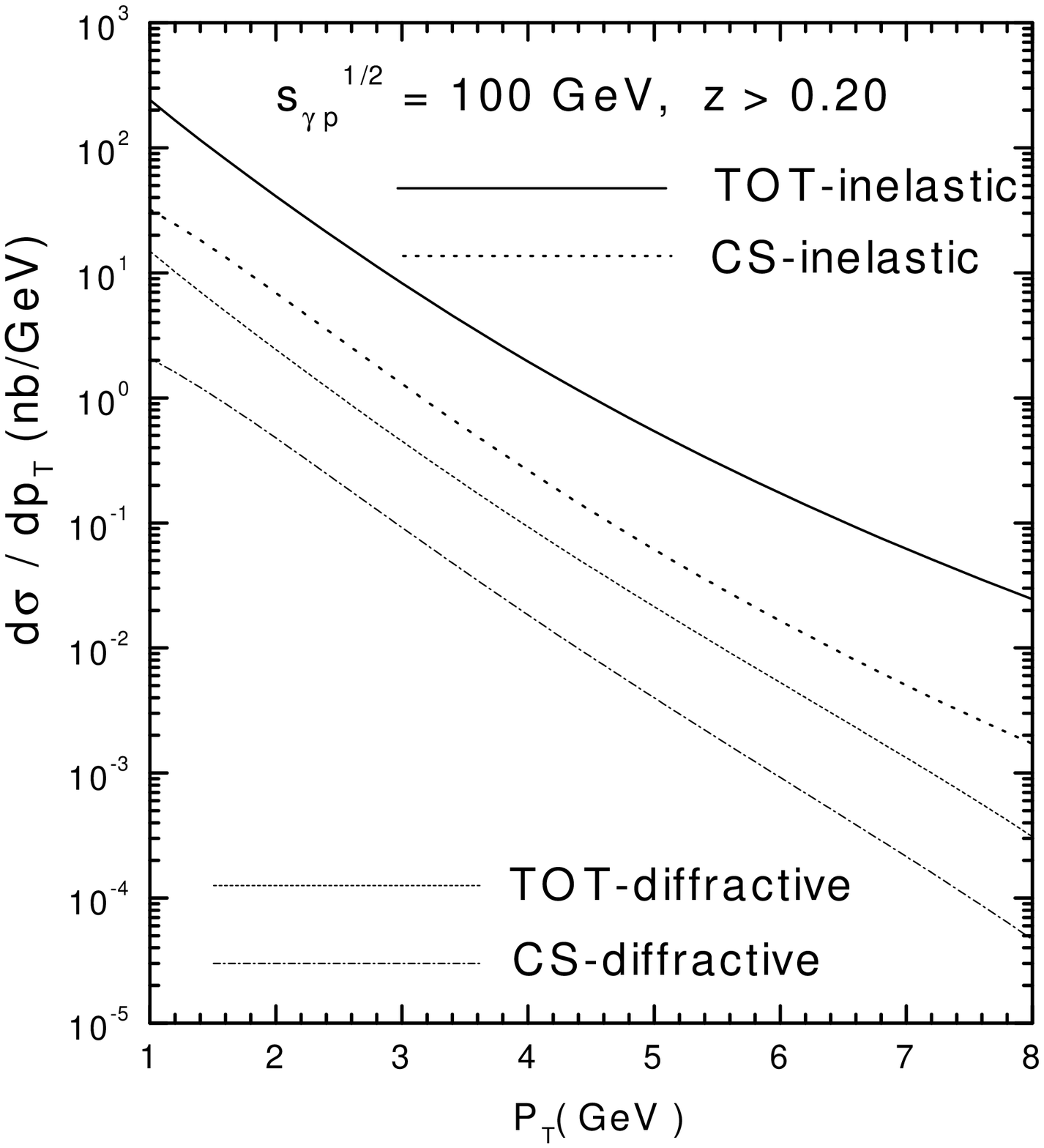,width=90mm}}
Fig.6. Transverse momentum ($P_T$) distribution $ d\sigma /dP_T $
for $\jpsi + jet$ production at HERA at
$\sqrt{s_{\gamma p}} = 100 {\rm GeV}$
with $z$ is integrated from  its lower cut $0.2$ to its kinematic maximum.
The thin dash-dotted and short dashed lines
represent the color-singlet cotribution and the sum of the color-singlet
and color-octet contributions to
$\jpsi + jet$ diffractive photoproduction $\gamma + p \to \jpsi + jet + p
+ X$ respectivly.
The thick solid and dotted lines curves are the sum of the color-singlet
and color-octet contributions and the color-singlet contribution to
inelastic $\jpsi + jet$ photoproduction $\gamma + p \to \jpsi + jet + X$
respectively.

\newpage
\centerline{\epsfig{figure=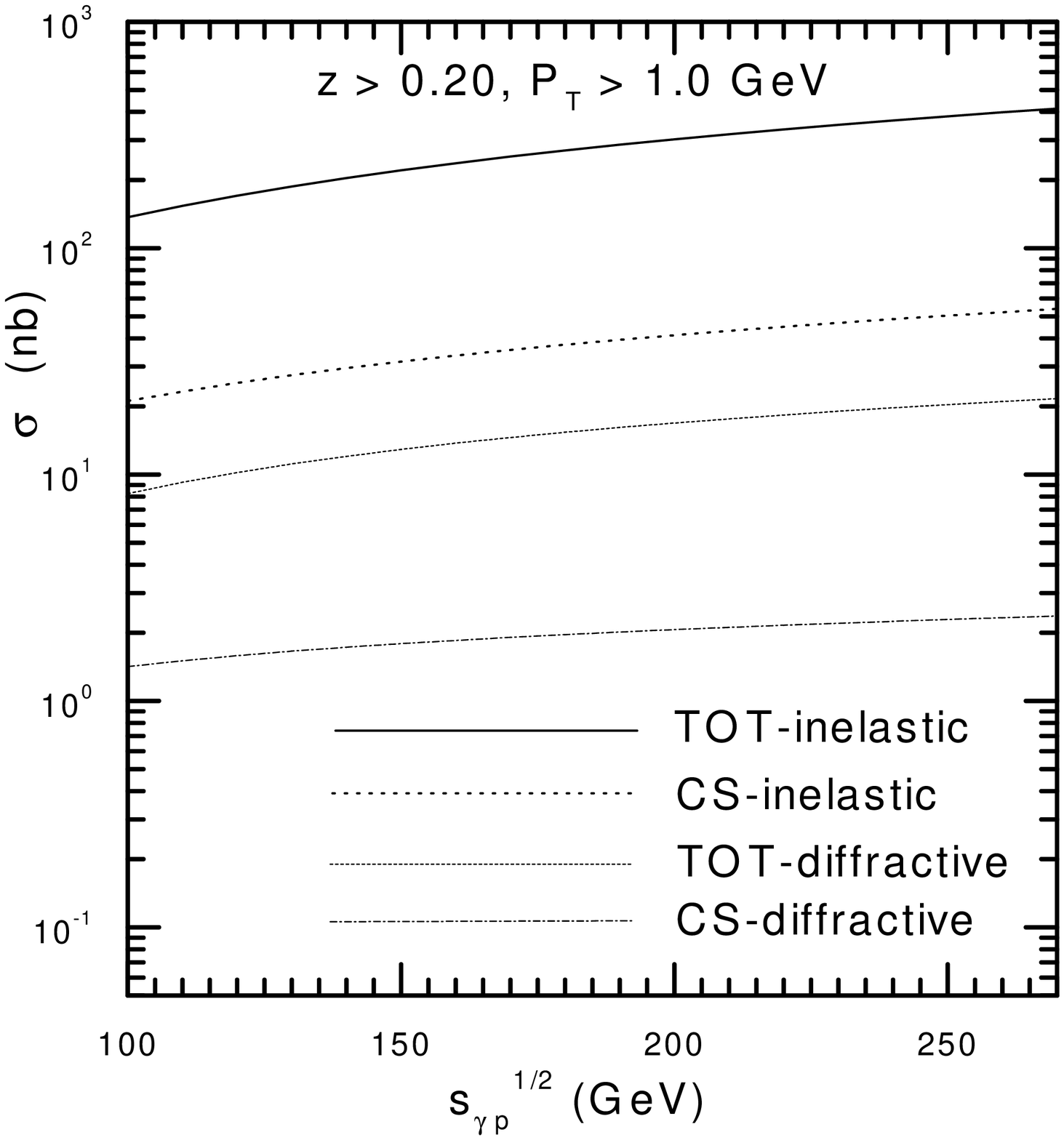,width=90mm}}
Fig.7. Total $\jpsi + jet$ photoproduction cross section for
$P_T \geq 1.0 {\rm GeV}, z\geq 0.2$ as a function of $\sqrt{s_{\gamma p}}$.
The thin dash-dotted and short dashed lines
represent the color-singlet cotribution and the sum of the color-singlet
and color-octet contributions to  $\jpsi + jet$ diffractive
photoproduction $\gamma + p \to \jpsi + jet + p + X$ respectivly.
The thick solid and dotted lines curves are the sum of the color-singlet
and color-octet contributions and the color-singlet contribution to
inelastic $\jpsi + jet$ photoproduction $\gamma + p \to \jpsi + jet + X$
respectively.
\end{figure}

\end{document}